\documentclass[prx,twocolumn,amsmath,amssymb]{revtex4}

\usepackage[dvipdfmx]{xcolor}
\usepackage{bm}
\usepackage{dcolumn}% Align table columns on decimal point
\usepackage[pdftex]{graphicx}    % for arXiv

\usepackage{slashed}
\usepackage{amsmath}\usepackage{accents}
\usepackage{ulem}

\newcommand{\cin}[1]{{\color{black}#1}}
\newcommand{\cinb}[1]{{\color{black}#1}}

\begin{document}

\title{
Higher-order topological heat conduction on a lattice for detection of corner states\\ 
%Detection of corner states via heat conduction on the breathing kagome lattice\\
%or\\
%Higher-order topological heat conduction on the breathing kagome lattice
}

\author{Takahiro Fukui$^1$, Tsuneya Yoshida$^2$, and Yasuhiro Hatsugai$^3$}
\affiliation{$^1$Department of Physics, Ibaraki University, Mito 310-8512, Japan}
\affiliation{$^2$Department of Physics, Kyoto University, Kyoto 606-8502, Japan}
\affiliation{$^3$Institute of Physics, University of Tsukuba, 1-1-1 Tennodai, Tsukuba, Ibaraki 305-8571, Japan}

\date{\today}

%%% abstract %%%
\begin{abstract}
A heat conduction equation on a lattice composed of nodes and bonds is formulated 
assuming the Fourier law and the energy conservation law. 
Based on this equation, 
we propose a higher-order topological heat conduction model on the breathing kagome lattice.
We show that the temperature measurement at a \cin{corner} node can detect the corner state
which causes rapid heat conduction toward the heat bath, 
and that several-nodes measurement can determine the precise energy of the corner states.
\end{abstract}

\pacs{
}

\maketitle

\section{Introduction}

The bulk-edge correspondence, which was originally proposed for the quantum Hall effect \cite{Hatsugai:1993fk,Hatsugai:1993aa},
plays a central role nowadays in various symmetry-protected topological (SPT) phases \cite{Kane:2005aa,Kane:2005ab,Qi:2008aa,Hasan:2010fk,Qi:2011kx}.  
Recent developments have revealed the universality of the bulk-edge correspondence,  including classical and/or 
nonequilibrium systems such as photonic crystals 
\cite{Raghu:2008vo,Haldane:2008to,Wang:2009aa,Ozawa:2019us,Kivshar:2019wn},
phononic systems
\cite{Prodan:2009vz,Savin:2010wo,Kane:2013aa,Kariyado:2015aa,Susstrunk:2015uo,Chien:2018tk,Yoshida:2019wg},
electrical circuits
\cite{Albert:2015wa,Lee:2018aa,Helbig:2020wq,Yoshida:2020vt},
hydrodynamics
\cite{Delplace:2017ui,Sone:2019vz}, etc.

Recently, \cin{inspired by the heat conduction in anti-PT symmetric systems \cite{doi:10.1126/science.aaw6259}}, heat conduction on a one-dimensional lattice 
which corresponds to a coarse discretization of the diffusion equation 
was theoretically proposed in Ref. \cite{Yoshida:2021vt}.
Incorporating alternating thermal diffusivities similar to the Su-Schrieffer-Heeger (SSH) model \cite{Su:1979aa},
they  showed \cite{Yoshida:2021vt} that 
edge states have a significant impact on heat conduction next to the heat bath.
\cin{Such an effect has} been experimentally 
observed  in \cin{Refs. \cite{https://doi.org/10.1002/adma.202202241,Hu:2022aa}}, indeed. 
It was also pointed out that the heat conduction on a lattice 
is reflected by the continuum diffusion equation with a similar SSH-like structure \cite{Makino:2022aa}.
This implies the usefulness of the heat conduction phenomena for experimental realization of
SPT phases. 

On the other hand, as new types of boundary states due to topological origin, the higher-order topological insulators (HOTI) 
\cite{Slager:2015aa,Hashimoto:2016aa,Kunst:2017aa,Benalcazar:2017ab,Benalcazar:2017aa,Hashimoto:2017aa,Kunst:2018aa,Schindler:2018ab,Ezawa:2018aa}
have been attracting much current interest. The boundary states called corner states, hinge states,  etc.,
inherent in the HOTI have been observed  in various systems 
such as electrical circuits \cite{Imhof:2018aa}, sonic crystals \cite{Zhang:2019ab}, and photonic crystals \cite{Ota:2019aa}.

In this paper, we propose a higher-order topological heat conduction system for
experimental detection of corner states. 
We demonstrate that the corner states cause 
rapid heat conduction at corners next to heat bath, and temperature measurement of several nodes enables to determine the 
precise energy of the corner states. This is due to the localization properties of the corner states.

This paper is organized as follows.
In the next section, we formulate a generic heat conduction on a lattice based on the Fourier law and 
the energy conservation. 
In Sec. \ref{s:kagome}, we apply it
to the heat conduction on the breathing kagome lattice, which is one of 
the typical models of the HOTI \cite{Ezawa:2018aa}.
In Sec. \ref{s:single_node}, we show that the measurement of the temperature only at \cinb{a} single corner is enough to detect 
the corner state. We also propose more precise measurement of the energy of the corner states using several 
nodes around the corner in Sec. \ref{s:effective}.
In Sec. \ref{s:sb},  the effect of the Stefan-Boltzmann thermal radiation was estimated.
In the Appendix, we argue the relationship between the lattice heat equation and the continuum heat equation, using the one-dimensional model
with a SSH-like structure.

\section{Heat conduction on a lattice}

\begin{figure}[h]
\begin{center}
\begin{tabular}{c}
\includegraphics[width=0.6\linewidth]{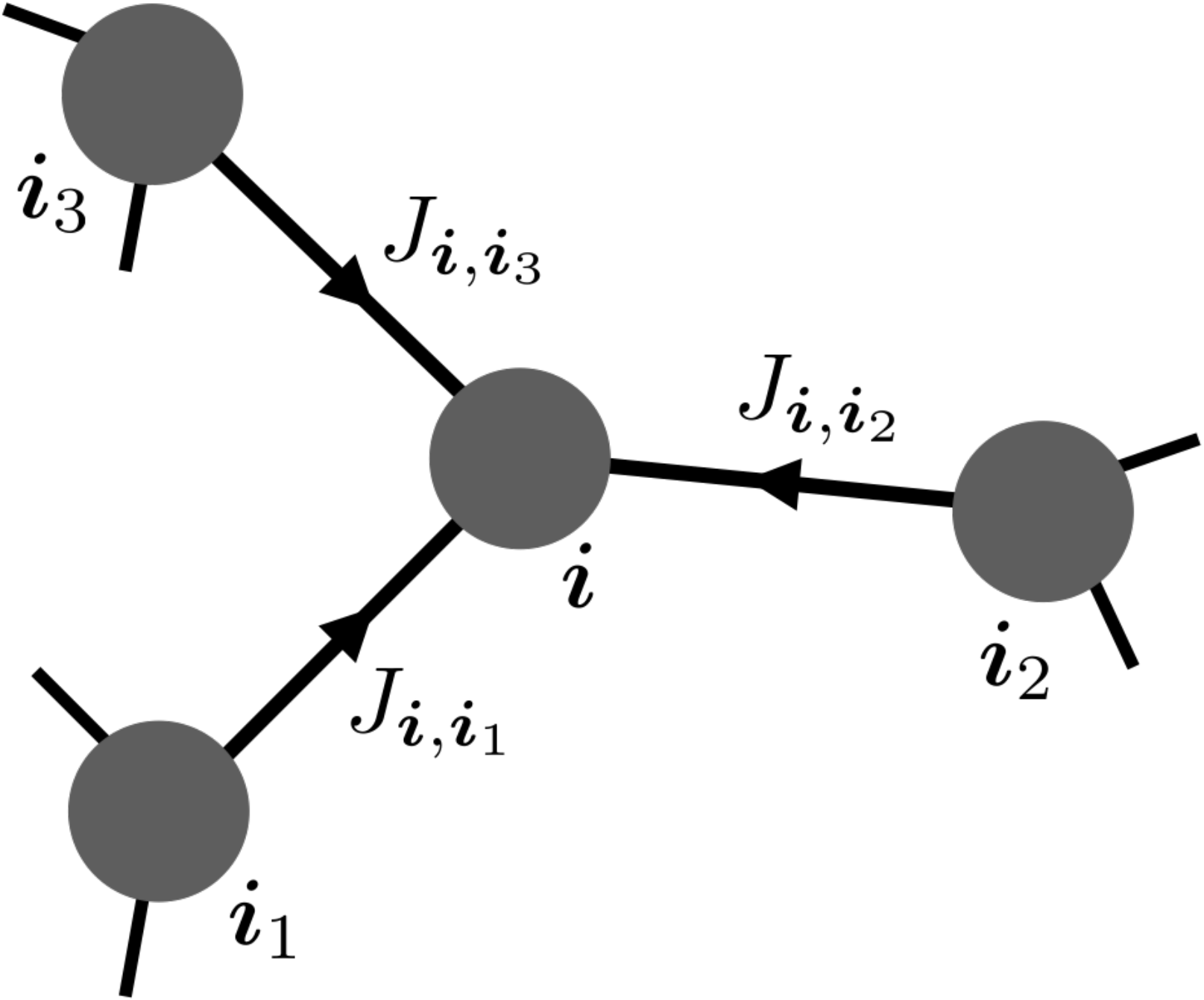}
\end{tabular}
\caption{
Nodes  specified by $\bm i=(i_1,i_2)$, and bonds between them. 
%$\hat \mu$  ($\mu=1,2$) is the unit vector toward $\mu$ direction.
}
\label{f:continuity}%-----------------------------------------------
\end{center}
\end{figure}

%Let us consider the Kagome lattice consisted of metallic nodes $\bm i$ with heat capacity $C$ 
%and two kinds of metallic bonds with thermal conductivities,  $K_1$ and $K_2$.
Let us consider heat conduction on lattice systems composed of nodes which are connected by bonds, 
as illustrated in Fig. \ref{f:continuity}.
We assume that the node at $\bm i$ has volume $V_{\bm i}$ and surface area $S_{\bm i}$ characterized by the specific heat capacity $c_{\bm i}$ 
and density $\rho_{\bm i}$,
whereas the bond between $\bm i$ and $\bm i'$ has the area of cross section $A_{\bm i,\bm i'}$ and the length $L_{\bm i,\bm i'}$ 
with the thermal conductivity $k_{\bm i,\bm i'}$.
We assume that the heat current flowing along the bond from $\bm i'$ and $\bm i$ is given by the 
Fourier law,
\begin{alignat}1
%J_{\bm i,\bm i'}=-K_{\bm i,\bm i'}(\phi_{\bm i}-\phi_{\bm i'}).
J_{\bm i,\bm i'}=-K_{\bm i,\bm i'}(T_{\bm i}-T_{\bm i'}),
\label{FouLaw}%---
\end{alignat} 
where 
%$\phi_{\bm i}$ 
$T_{\bm i}$ 
is the temperature of nodes at $\bm i$ 
%with respect to the temperature of the heat bath $\phi_0$, 
and $K_{\bm i,\bm i'}$ is the thermal conductance
given, using the thermal conductivity $k_{\bm i,\bm i'}$, by
\begin{alignat}1
K_{\bm i,\bm i'}=k_{\bm i,\bm i'}\frac{A_{\bm i,\bm i'}}{L_{\bm i,\bm i'}}.
\end{alignat}
At each node $\bm i$ with the heat capacity $c_{\bm i}\rho_{\bm i} V_{\bm i}$,
we require the continuity equation, 
\begin{alignat}1
%c_{\bm i}\rho_{\bm i} V_{\bm i}\partial_t\phi_{\bm i}
c_{\bm i}\rho_{\bm i} V_{\bm i}\partial_tT_{\bm i}
=\sum_nJ_{\bm i,\bm i_n}+\cin{q_{\bm i}},
\label{Continuity}%---
\end{alignat}
where $n$ of $\bm i_n$ labels all the bonds connected with the node $\bm i$ (see Fig. \ref{f:continuity}), and
%$n$ labels all the bonds connected with the node $\bm i$, and
%$\partial_\mu^* f_{\bm i}\equiv f_{\bm i}-f_{\bm i-\mu}$ stands for the backward difference,
\cin{$q_{\bm i}$} denotes some source term of heat.
In this paper, we consider effects of the thermal radiation obeying the Stefan-Boltzmann law, 
$\cin{q_{\bm i}}=\varepsilon\sigma S_{\bm i}(T_0^4- T_{\bm i}^4)$, where $T_0$ is the temperature of the heat bath,
$\sigma$ is the Stefan-Boltzmann constant, and $\varepsilon$ is emissivity $0\le\varepsilon\le1$.
Here, we assume that the thermal radiation occurs at the nodes only.
Together with the Fourier law, the continuity equation can be written as
\begin{alignat}1
%\partial_t\phi_{\bm i}&=-\sum_{n}D_{{\bm i},{\bm i}_n}
%(\phi_{{\bm i}}-\phi_{{\bm i}_n})-\sigma S_{\bm i}(\phi_0^4- \phi_{\bm i}^4)
\partial_tT_{\bm i}&=-\sum_{n}D_{{\bm i},{\bm i}_n}
(T_{{\bm i}}-T_{{\bm i}_n})+\sigma_{\bm i} (T_0^4- T_{\bm i}^4)
\nonumber\\
&
%\equiv-\sum_{\bm j}{\cal H}_{\bm i\bm j}\phi_{\bm j}
\equiv-\sum_{\bm j}{\cal H}_{\bm i\bm j}T_{\bm j}
%+\sigma S_{\bm i}(\phi_0^4- \phi_{\bm i}^4),
+\sigma_{\bm i}(T_0^4- T_{\bm i}^4),
\label{GenDifEqu}%---
\end{alignat} 
where 
%$n$ of $\bm i_n$ labels all the bonds connected with the node $\bm i$  and 
\begin{alignat}1
D_{\bm i,\bm i'}&\equiv \frac{k_{\bm i,{\bm i}'}}{c_{\bm i}\rho_{\bm i}}\frac{A_{\bm i,{\bm i}'}}{V_{\bm i}L_{\bm i,{\bm i}'}},
\nonumber\\
\sigma_{\bm i}&\equiv \frac{\varepsilon\sigma}{c_{\bm i}\rho_{\bm i}}\frac{S_{\bm i}}{V_{\bm i}}.
\label{GenDifCon}%---
\end{alignat}
Equation (\ref{GenDifEqu}) with  (\ref{GenDifCon}) 
is the heat conduction equation on a lattice we study in this paper.
In Eq. (\ref{GenDifCon}), $k_{\bm i,{\bm i}_n}/(c_{\bm i}\rho_{\bm i})$ seems the conventional thermal diffusivity.
%, which is nothing but the 
%diffusion constant characterizing the continuum heat conduction equation. 
However, note that
%On the other hand, in the lattice model, 
the thermal conductivity $k_{\bm i,\bm i'}$ is
%depends on the geometrical structure of the lattice system, and is defined by  the thermal conductance 
that of the bonds, whereas the heat capacity  is
$c_{\bm i}\rho_{\bm i}$
%, and the heat capacity 
that of the nodes.
Moreover, the thermal conductance $D_{\bm i,\bm i'}$ depends on the geometrical structure of the lattice system.
%, $c_{\bm i}\rho_{\bm i}V_{\bm i}$.
%, whereas 
%in the continuum diffusion equation, the diffusion constant should be given by the local thermal conductance and heat capacity.
Therefore, in the diffusion on the lattice, diffusion constants should be regarded just as effective parameters, which may be 
obtained by suitable renormalization from the continuum heat conduction equation. 
We exemplify this fact in the Appendix, comparing the heat equation in one dimension 
discretized by fine meshes and coarse meshes.

As an experimental implementation, we assume that initially all nodes have the same temperature as the heat bath $T_0$, 
and that at time 
$t=0$,  some nodes are heated up to $T_0+\delta T$, which causes the heat conduction on a lattice.
%and $t>0$ the change of temperature of each node is observed.
In such an experiment, it may be convenient to 
%For the local temperature $T_{\bm i}$, let us 
introduce the notation
\begin{alignat}1
T_{\bm i}= \delta T(\phi_{\bm i}+\phi_0),\quad (0\le\phi_{\bm i}\le1),
\end{alignat}
where 
%\csout{$\delta T+T_0$ is the initial highest temperature and} 
$\phi_0=T_0/\delta T$.
Then Eq. (\ref{GenDifEqu}) can be written as
\begin{alignat}1
\partial_t\phi_{\bm i}&
%=-\sum_{n}D_{{\bm i},{\bm i}_n}
%(\phi_{{\bm i}}-\phi_{{\bm i}_n})-\sigma S_{\bm i}(\phi_0^4- \phi_{\bm i}^4)
%\nonumber\\
%&
=-\sum_{\bm j}{\cal H}_{\bm i\bm j}\phi_{\bm j}
+\alpha_{\bm i}\left[\phi_0^4- (\phi_{\bm i}+\phi_0)^4\right],
\label{GenDifEqu2}%---
\end{alignat} 
where 
\begin{alignat}1
\alpha_{\bm i}=\sigma_{\bm i}\delta T^3.
\end{alignat}

In the next section, we will mainly consider the model without thermal radiation, $\sigma=0$,
\cin{since it will turn out that this effect is very limited, as shown in Sec. \ref{s:sb}.} 
In this case, it may be convenient to introduce 
%Let us solve 
the eigenvalue equation:
\begin{alignat}1
\sum_{\bm j}{\cal H}_{\bm i\bm j}\phi_{\bm jn}=\phi_{\bm in}\varepsilon_n.
\label{Ham1}%---
\end{alignat}
For simplicity, let us introduce the vector-notation $\bm\phi_n$ which is  a vector whose components are $\phi_{\bm in}$. 
When $\sigma=0$, the time evolution of 
the local temperature $\phi_{\bm i} (t)$ starting from the initial temperature distribution
%\csout{
%$\bm\phi_{\rm i}$ which is also the same vector notation as introduced above whose components are} 
$\phi_{\bm i}(0)$, 
\cinb{which is also denoted as a vector $\bm \phi(t)$ or $\bm \phi(0)$}, 
%\equiv (\phi_{\bm i_1}(0),\phi_{\bm i_2}(0),\cdots)$ 
is simply obtained as
\cinb{
\begin{alignat}1
\bm\phi(t)&=e^{-{\cal H}t}\bm\phi(0)
%\nonumber\\&
=\sum_ne^{-\varepsilon_nt}\bm\phi_{n}\bm\phi_{n}^T\bm \phi(0),
\label{TimEvo}%---
\end{alignat}
}
where \cinb{$\bm\phi_{n}^T\bm\phi(0)=\sum_{\bm j}\phi_{\bm j n}\phi_{\bm j}(0)$.}

\section{Heat conduction on the breathing kagome lattice}\label{s:kagome}%---

\begin{figure}[h]
\begin{center}
\begin{tabular}{c}
\includegraphics[width=0.8\linewidth]{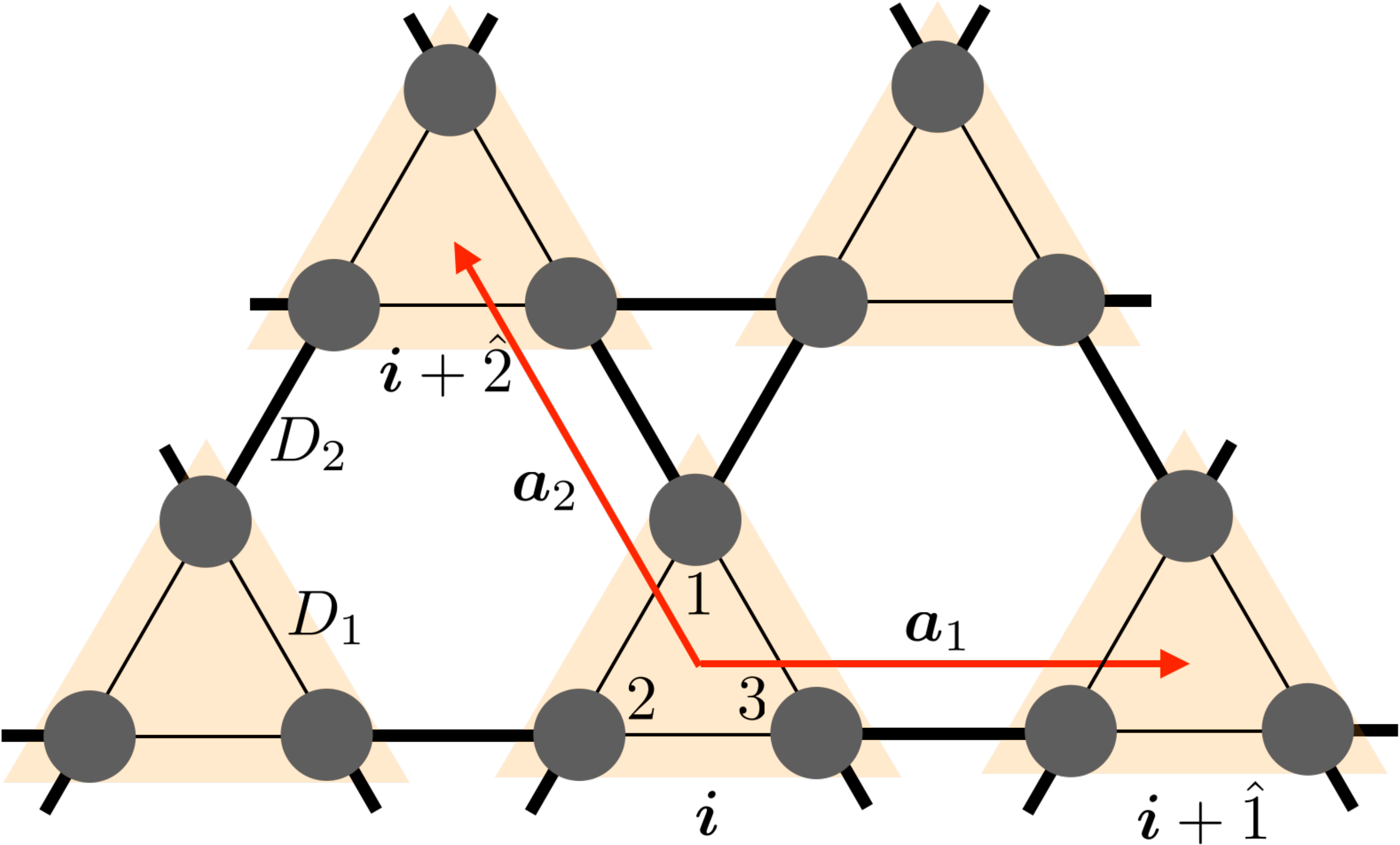}
\end{tabular}
\caption{
Nodes and bonds forming the kagome lattice. The orange triangle indicates the unit cell, and the red arrows are 
primitive translation vectors.
}
\label{f:lattice}%-----------------------------------------------
\end{center}
\end{figure}

The breathing kagome lattice is one of the well-known examples of the HOTI \cite{Ezawa:2018aa}
\cin{(See also Ref. \cite{PhysRevB.106.085420}, which examined corner states of the kagome lattice in proximity to superconductors)}.
Therefore, for the detection of the corner states, 
let us consider heat conduction on the breathing kagome lattice illustrated in Fig. \ref{f:lattice}, where 
we have defined the unit cell composed by three nodes specified by $a=1,2,3$.
The primitive translation vectors are $\bm a_1$ and $\bm a_2$, and, in addition, 
$\bm a_3\equiv -(\bm a_1+\bm a_2)$ is introduced,  for convenience.
These vectors have the same length $|\bm a_j|\equiv a$.
We assume that all nodes are identical, whereas the bonds have two kinds of geometrical forms characterized by 
$A_j/L_j$ $(j=1,2)$, 
or two kinds of thermal conductivities, $k_j$ $(j=1,2)$, both of which yield two kinds of thermal diffusion constants $D_j$
($j=1,2$).
It follows from Eqs. (\ref{GenDifEqu2}) and (\ref{GenDifCon}) that the heat equation on the breathing kagome lattice
is given by
\begin{alignat}1
\partial_t\phi_{a,\bm i}(t)&=\sum_{b(\ne a)}
\left[D_1(\phi_{b,\bm i}-\phi_{a,\bm i})-D_2(\phi_{a,\bm i}-\phi_{b,\bm i-\epsilon^{abc}\bm a_c})
\right]
\nonumber\\
&+\alpha(\phi_0^4-\phi_{a,\bm i}^4),
\label{HeaEquKag}%---
\end{alignat}
where 
\begin{alignat}1
D_j&\equiv 
\left\{
\begin{array}{ll}
\displaystyle 
\frac{k}{c\rho}\frac{A_j}{VL_j}\equiv \frac{d}{l_j^2} & \\
\mbox{or } &\quad (j=1,2),\\
\displaystyle 
\frac{k_j}{c\rho}\frac{A}{VL}\equiv \frac{d_j}{l^2} &\\
\end{array}
\right.
\nonumber\\
\alpha&=\frac{\varepsilon\sigma}{c\rho}\frac{S}{V}\delta T^3.
\label{ParKag1}%---
\end{alignat} 
In this equation, we assume that $d$ or $d_j$ with some suitable renormalization
\cinb{denotes} the conventional thermal \cinb{diffusivity} for the continuum system, and
the parameter $l_j$ with the dimension of length characterizes the length scale of the present lattice system.
Since we treat the nodes and bonds as simple points and lines, ignoring their geometric shapes, 
the model  includes  only the lattice constant $a$ which has the dimension of length.  
In what follows, we introduce the parametrization 
\begin{alignat}1
D_1\equiv \frac{d_1}{a^2},\quad D_2\equiv \frac{d_2}{a^2},\quad (D_1=\delta D_2)
\label{ParKag2}%---
\end{alignat}
using the lattice constant $a$ of the kagome lattice. 
\cin{Here, the parameter $\delta$ characterizes the topological phases of the breathing kagome lattice.}
Let us also define a typical time scale
\begin{alignat}1
\tau=\frac{1}{D}=\frac{a^2}{d},
\end{alignat}
where $D\equiv \min (D_1,D_2)$ and $d\equiv \min (d_1,d_2)$.

\begin{figure}[h]
\begin{center}
\begin{tabular}{c}
\includegraphics[width=0.7\linewidth]{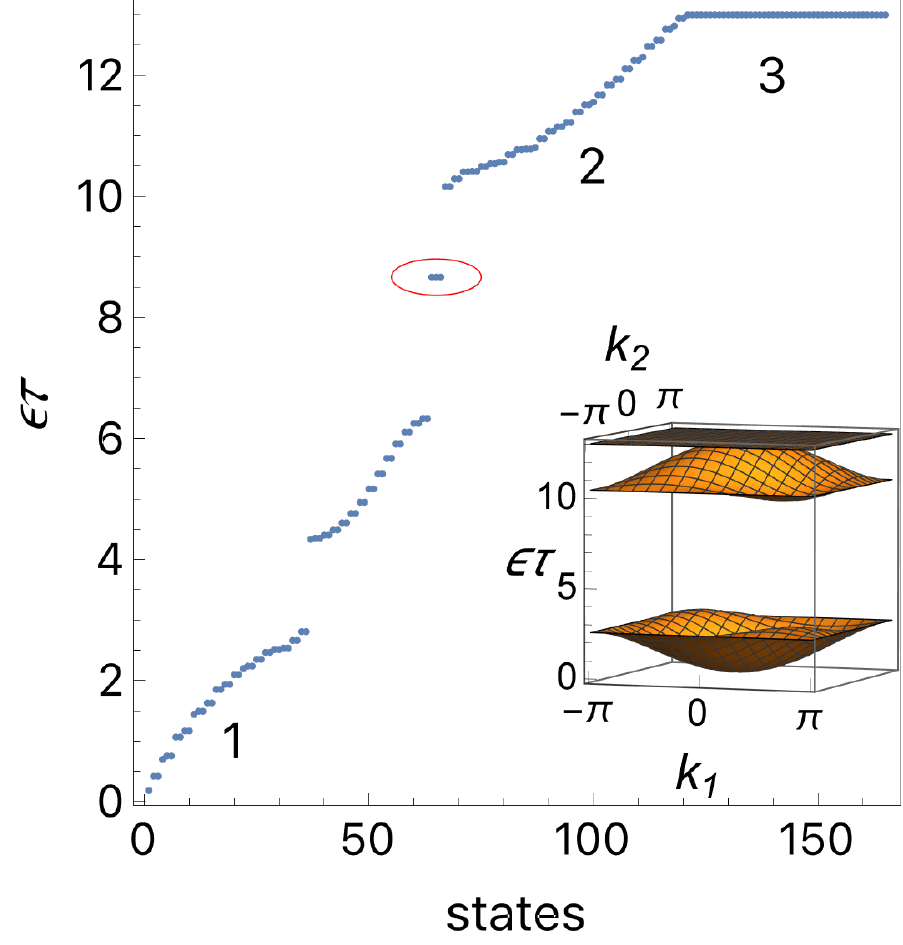}
\end{tabular}
\caption{
Spectrum of $\cal H$ in Eq. (\ref{Ham1}) for the breathing kagome lattice system (\ref{HeaEquKag}) at $\delta=0.3$ of equilateral triangle 
with 10 unit cells per side.
%, \cin{where $\delta$ is defined in Eq. (\ref{ParKag2}).}
The threefold degenerate states encircled are corner states, other bands specified by 
numbers one, two, and three are the bulk bands, and the band with no number is due to the edge states.
The inset shows the bulk spectrum.
%in Eq. (\ref{Ham1}) for the breathing kagome lattice system of equilateral triangle with 10 unit cells per side.The red dots indicate the %corner states.
}
\label{f:full}%-----------------------------------------------
\end{center}
\end{figure}

\begin{figure}[h]
\begin{center}
\begin{tabular}{cc}
\includegraphics[width=0.49\linewidth]{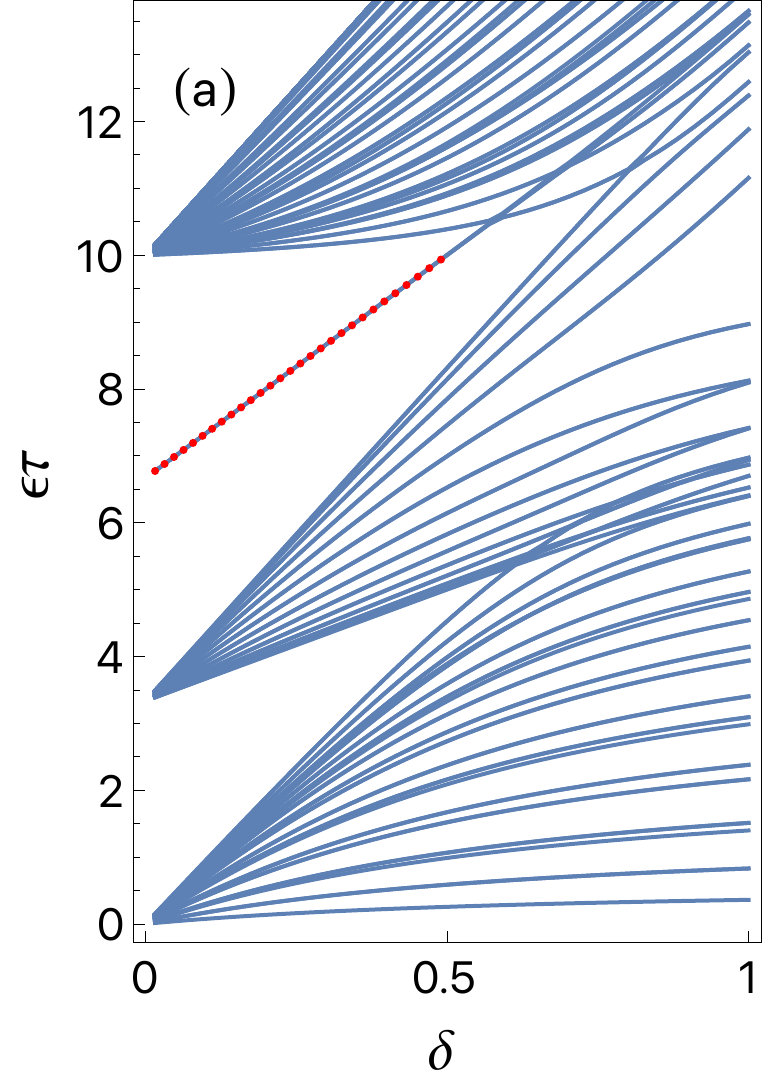}
&\includegraphics[width=0.49\linewidth]{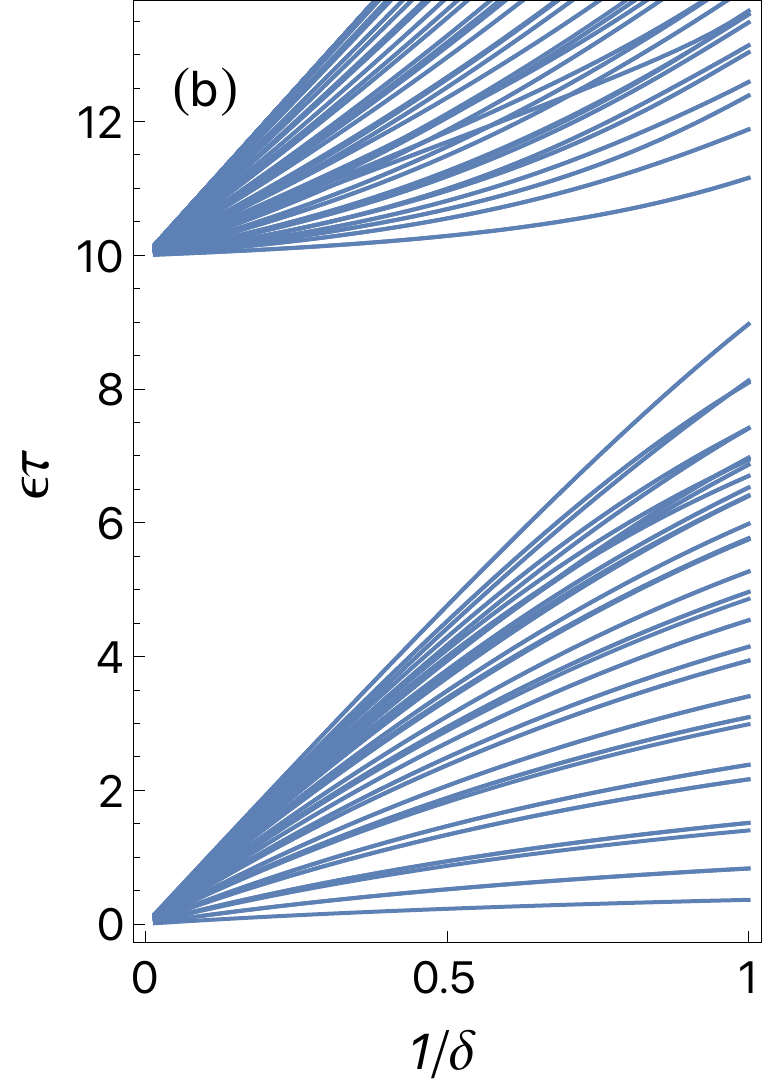}
\end{tabular}
\caption{
The spectrum of the same equilateral triangle system of Fig. \ref{f:full}  as a function of (a) $\delta$ and (b) $1/\delta$,
\cin{where $\delta$ is defined in Eq. (\ref{ParKag2}).}
The red dots indicate the corner states.
}
\label{f:corner}%-----------------------------------------------
\end{center}
\end{figure}

\subsection{Spectrum of a finite system}

This model show three bands, lowest and middle dispersive bands with a gap 
as well as a flat band at the top of the middle band (inset of Fig. \ref{f:full}).
To investigate the corner states, 
let us consider the system of the triangle shape surrounded by heat bath, including $N$ nodes.
For example, $N$ is given by $N=55\times3=165$ for the triangle with ten unit cells per side. 
Figure \ref{f:full} shows the spectrum for $\delta=0.3$ in the HOTI phase, \cin{where $\delta$ is defined in Eq. (\ref{ParKag2}).}
We can see  two dispersive bulk bands and a flat band, as well as 
threefold degenerate corner states and edge states in between the gap of dispersive bands one and two.
We \cinb{show} in Fig. \ref{f:corner}, the spectrum  of $\cal H$ as a function of $\delta$.
Since this model belongs to HOTI phase for $0<\delta <1/2$, it \cinb{exhibits} the corner states in this range
\cite{Ezawa:2018aa}. 
Even for $1/2<\delta<1$, there exist corner states, but embedded in the bulk upper band.
For $1<\delta$, the model belongs to a trivial phase without corner  states as shown in Fig. \ref{f:corner}(b). 
Not only corner states but also edge states can be seen in Fig. \ref{f:corner}(a), which are missing in the trivial regime 
in Fig. \ref{f:corner} (b).

\begin{figure}[h]
\begin{center}
\begin{tabular}{c}
\includegraphics[width=0.8\linewidth]{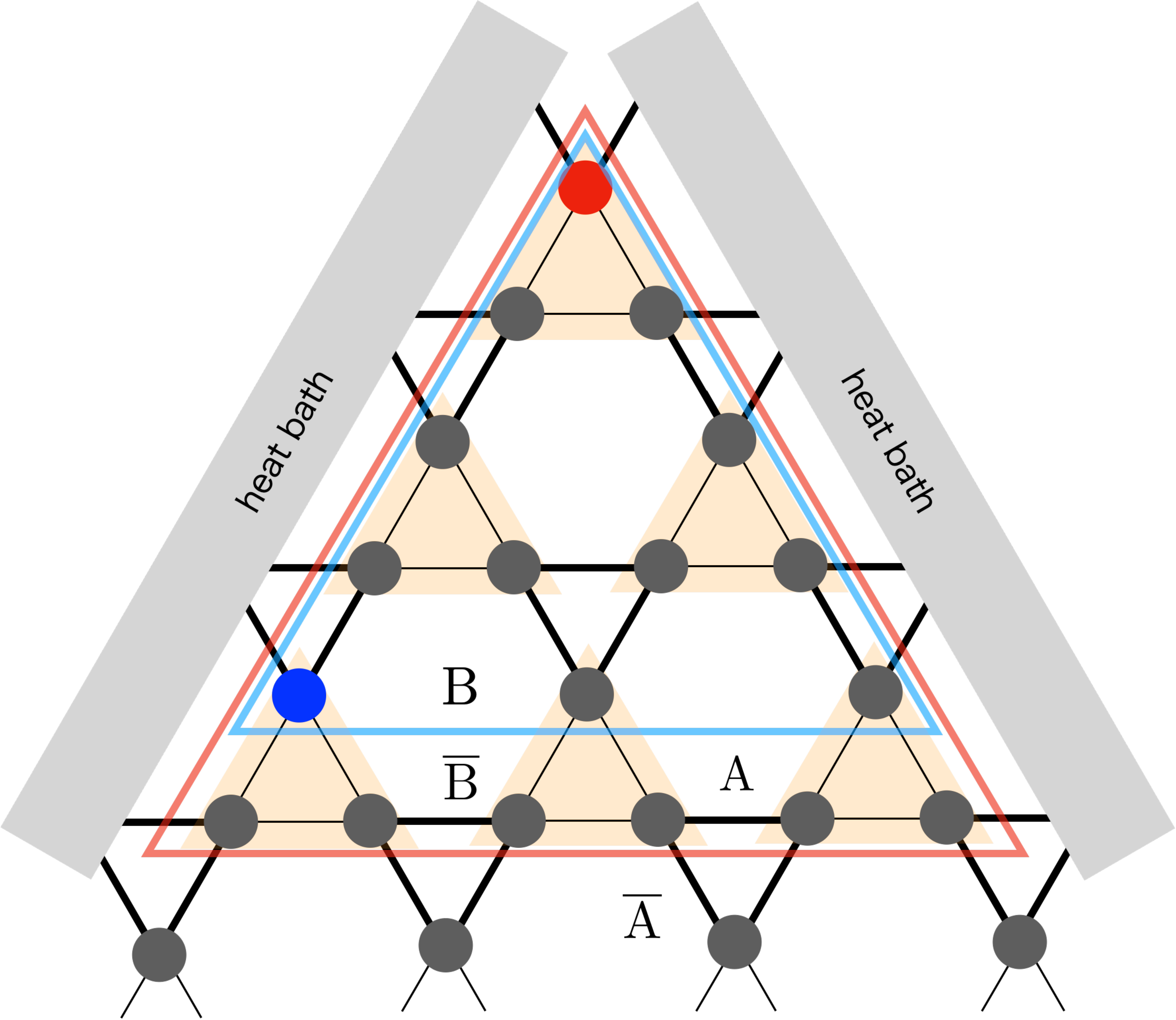}
\end{tabular}
\caption{
Top corner. The red and blue nodes are the nodes used for single-node measurement in Sec. \ref{s:single_node}.
The nodes surrounded by red (A) and blue (B) triangles are used for deriving effective 
Hamiltonians. Regions A and B include 18 and 12 nodes, respectively, in Sec. \ref{s:effective}.
}
\label{f:ini}%-----------------------------------------------
\end{center}
\end{figure}

\subsection{Toward the detection of the corner states}

Toward experimental detection of the corner states in the heat conduction system, 
we first consider the temperature measurement  at one corner node \cite{Yoshida:2021vt}, 
which indeed reveals the existence of the corner states.
%, in which we argue that
% the corner state  can be detected just from one node measurement of the temperature.
Next, we discuss an improved way to determine the corner states by using data of temperature of several nodes around a corner
which provides more accurate energy of the corner state.
This may be called an effective Hamiltonian approach.
For simplicity, we consider the case of $\alpha=0$.

\subsubsection{Heat conduction at a corner}\label{s:single_node}

\begin{figure}[h]
\begin{center}
\begin{tabular}{cc}
\includegraphics[width=0.5\linewidth]{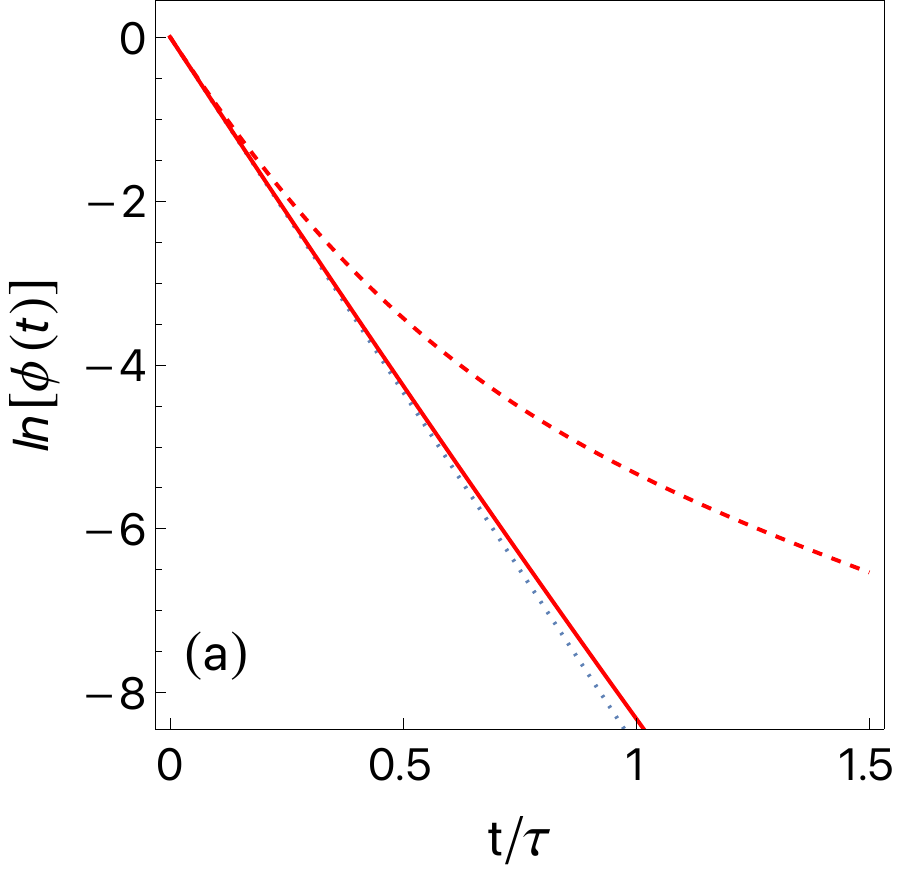}
&\includegraphics[width=0.5\linewidth]{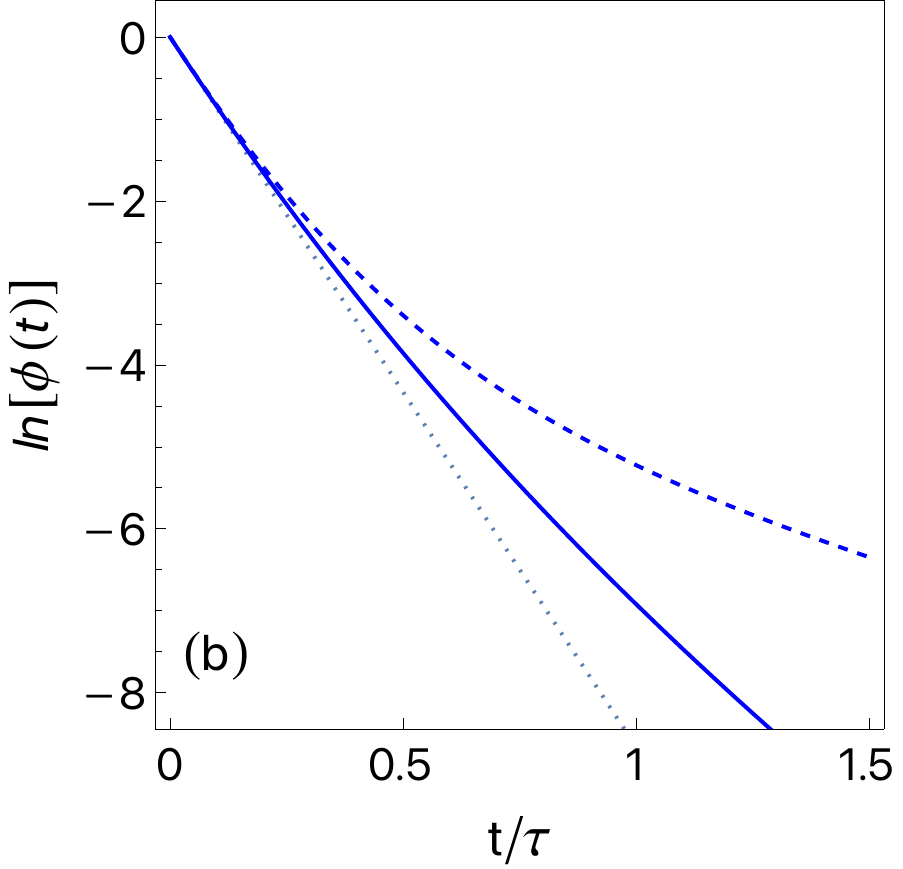}\\
\includegraphics[width=0.5\linewidth]{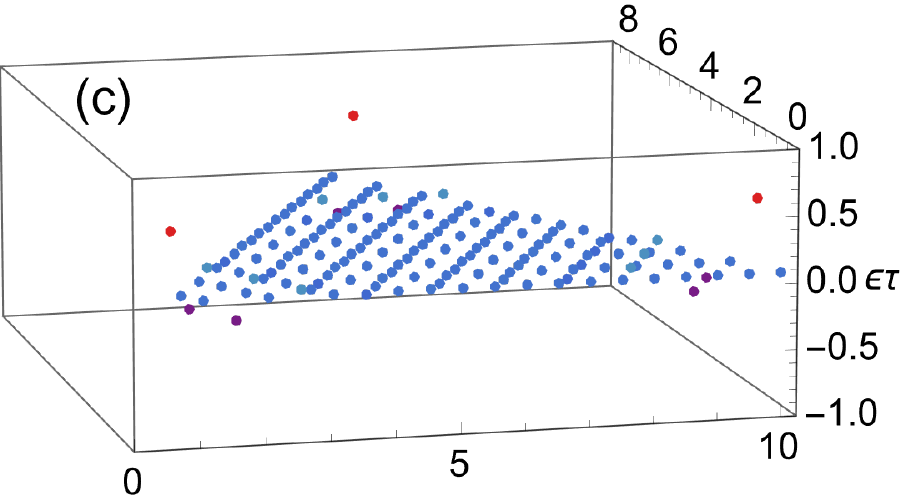}
&\includegraphics[width=0.5\linewidth]{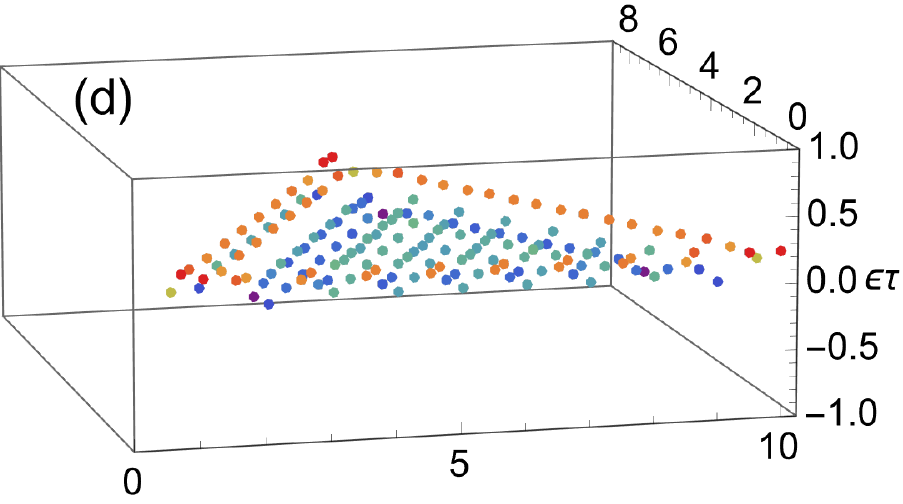}
\end{tabular}
\caption{
Single-node temperature as a function of $t/\tau$. (a) and (b) are the corner node and edge node highlighted in red and blue
in Fig. \ref{f:ini}, respectively. Full curves and dashed curves are the cases $\delta=0.3$ and $\delta=1/0.3$, respectively, 
which belong to the topological and trivial phases. Thin dotted lines are $-\varepsilon_{\rm c}t$, where $\varepsilon_{\rm c}$ 
is the energy of the corner states $\varepsilon_{\rm c}\tau=2(D_1+D_2)/D$.
\cin{(c) and (d) show the examples of the temperature profiles on the lattice for the corner states (No. 66) and edge states 
(No. 37, the lowest of the edge band), respectively, in the topological phase.}
}
\label{f:ces}%-----------------------------------------------
\end{center}
\end{figure}

Assume that all the nodes are at the room temperature, $\phi_{\bm i}=0$. Let us give temperature $\phi_{\rm c}=1$ 
only at the corner node highlighted in red in Fig. \ref{f:ini} at time $t=0$, and measure how the given temperature decreases \cite{Yoshida:2021vt}. 
The result is shown in Fig. \ref{f:ces}(a), in which one can see a clear distinction between the topological phase (solid curve) 
and trivial phase (dashed curve) in a short time regime. In particular, the heat conduction at the corner node is dominated
in the topological phase by the corner state up to $t\sim\tau$.  
In passing, we point out that 
away from the corner but on the boundary highlighted by blue in Fig. \ref{f:ini}, 
edge states can also be  observed, as shown in Fig. \ref{f:ces}(b).
Indeed, the solid curve in the topological phase has clear difference from the dashed curve in the trivial phase.
Its curvature also suggests the dispersive band for the edge states.
\cin{For comparison, we show in Figs. \ref{f:ces}(c) and \ref{f:ces}(d) the temperature profiles on the lattice for the corner state and edge state, respectively.}

\subsubsection{Effective Hamiltonian approach}\label{s:effective}%---

So far we have shown that only one-node measurement of temperature is enough to detect 
the corner state in the heat conduction system. However, more precise determination of the energy of the corner state
is achieved if one measures the temperature at more than one node \cite{Yoshida:2021vt}.  
To this end, let us prepare several initial distributions distinguished by index $p$. 
As in Eq. (\ref{Ham1}), we introduce the vector notation $\bm\phi_p$
whose components are $\phi_{\bm ip}(0)$.
We assume that they are independent in the sense
\begin{alignat}1
\bm\phi_p^T\bm\phi_q=\delta_{pq}.
\label{Ort}%---
\end{alignat}
Since the local temperature $\phi_{\bm i}(t)$ depends on the initial distribution $\bm \phi_p$, let us denote
$\phi_{\bm ip}(t)$, for which we also use the vector notation $\bm \phi_p(t)$.
Then, the time evolution (\ref{TimEvo}) is written as
\begin{alignat}1
\bm\phi_p(t)=\sum_ne^{-\varepsilon_nt}\bm\phi_n\bm\phi^T_n\bm\phi_p.
\end{alignat}
%using the vector notation.
Taking the inner product with $\bm\phi_q$, we have \cite{Yoshida:2021vt}
\begin{alignat}1
{\cal T}_{pq}\equiv \bm\phi_q^T \bm\phi_p(t)=\bm\phi_q^Te^{-{\cal H}t}\bm\phi_p\equiv e^{-{\cal H}_{qp}t}.
\label{CalT}%---
\end{alignat}
The left hand side can be measured experimentally, whereas the right hand side tells how to compute the 
energy of the Hamiltonian $\cal H$. Namely, let $E_n$ be eigenvalues of the matrix $\cal T$. 
Then, the eigenvalues of $\cal H$ are given by $\varepsilon_n=-(\log E_n)/t$.
For a finite system with $N$ nodes,
if we prepare $N$ independent initial temperature distributions and measure the temperature 
at all $N$ nodes, we can reproduce the exact energies shown in Fig. \ref{f:full}, as carried out in  \cite{Yoshida:2021vt}.
%On the other hand, if we prepare only one independent initial distribution, we have the result in Sec. \ref{s:single_node}.

\begin{figure}[h]
\begin{center}
\begin{tabular}{cc}
\includegraphics[width=0.5\linewidth]{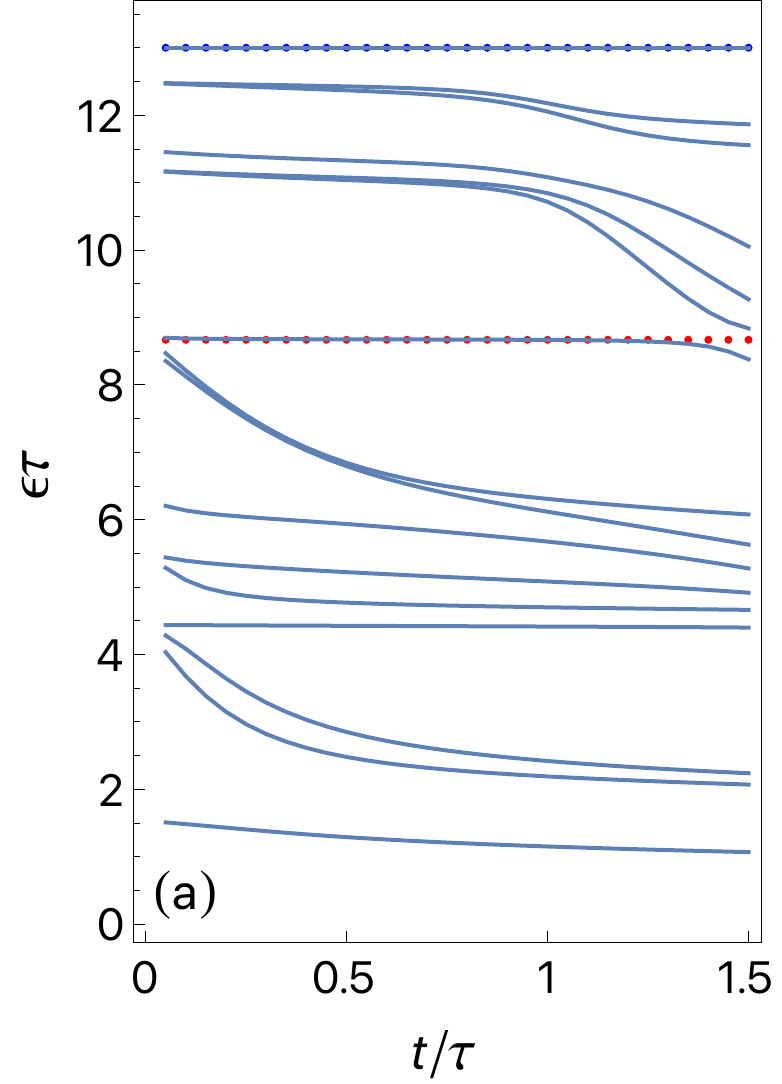}
&\includegraphics[width=0.5\linewidth]{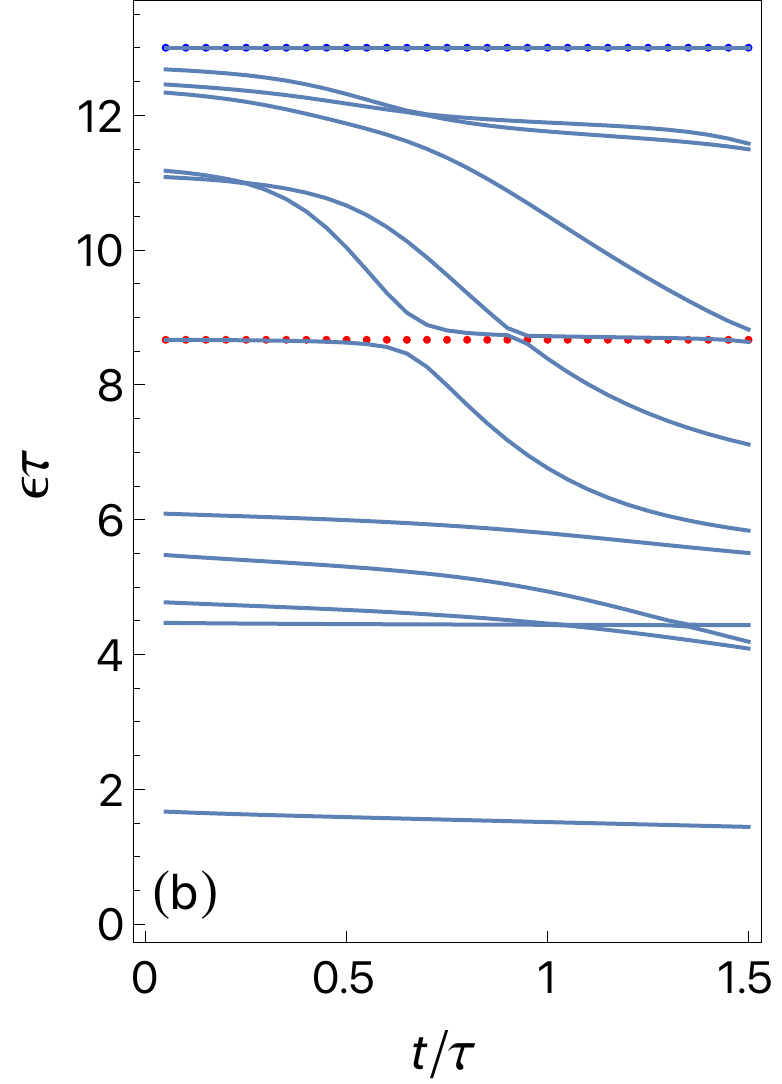}\\
\end{tabular}
\caption{
Spectra of effective Hamiltonian with restricted degrees of freedom as  functions of time. 
(a) and (b) correspond to region A and B in Fig. \ref{f:ini}.
Red and blue dots denote the exact energy of the corner state and the flat band, respectively. 
}
\label{f:eff}%-----------------------------------------------
\end{center}
\end{figure}

However, for the purpose of observing corner states and determining their energy, 
fewer initial conditions and fewer node temperature measurements are sufficient since each corner state is localized
around each corner. 
Actually, in Sec. \ref{s:single_node}, only one-node measurement at one corner with only one initial distribution 
enables to detect the existence of the corner mode. 
For such restricted measurements using reduced degrees of freedom, the energies, which 
should be constant along the time evolution, inevitably depend on time.
Nevertheless, the energy of the corner states is almost constant. This is due to their localized nature.
Thus, the time dependence of the energies could distinguish the energy of the corner state 
from those of other extended states.

For simplicity, we propose the following initial temperature distributions: 
We keep the temperature of all the nodes $\phi_{\bm j}=0$
except for that of node $\bm i$ which is set $\phi_{\bm i}=1$.
The number of such initial distributions is of course $N$ and they satisfy Eq. (\ref{Ort}). 
Here, we propose the restricted number of initial distributions to the nodes specified by triangles
%illustrated 
in Fig. \ref{f:ini}.
The whole system has 165 nodes, whereas 
the regions A and B include only 18 and 12 nodes, each of which is given the initial temperature $1$. 
Then, the measurements of the local temperature of each node in regions A and B at time $t$ allow us to calculate
the energies of the reduced Hamiltonian. 
In Fig. \ref{f:eff}, we show the energies of such a reduced Hamiltonian.

From the single-node measurement in Sec. \ref{s:single_node}, 
we expect that the energies of the effective reduced Hamiltonian are decreasing as functions of time, since
long-time behavior is controlled by low energy states, some of which are neglected in the reduced Hamiltonian.
%Remember that the energies of the total system shown in Fig. \ref{f:full} describes 
%how the energies are decreasing toward the heat bath. 
%When we focus on the restricted region A or B, the energies are flowing not only toward the heat bath but also out of A or B 
%regions, which are denoted as $\overline{\rm A}$ or  $\overline{\rm B}$. 
%This implies that the energies calculated by using the restricted region A or B, which should describes how the energies
%are flowing to the heat bath, decrease as a function of time.
In Fig. \ref{f:eff}, we plot the spectra of the reduce Hamiltonian for the restricted nodes in the A and B regions in Fig. \ref{f:ini}.
Indeed, Fig. \ref{f:eff} shows that all the energies are basically decreasing function. 
Among them,  the energies which are nearly degenerate with the corner state at $t\sim0$
are largely decreasing, as can be seen in Fig. \ref{f:eff}(a).
These are due to the states at the artificial boundary between A and $\overline{\mathrm{A}}$, which are connected by the large $D$ bonds.
On the other hand, almost constant energies are due to the states which are not affected by the artificial boundary.
The corner state may be the most typical state with such a property, because it has a  localized temperature profile. 
Another typical state may be the flat band: Huge degeneracy enables to create a localized state by suitable linear combination of degenerate states, as indicated by blue dots in Fig. \ref{f:eff}.

%Clearly, the corner state can be observed as a state with an almost constant energy even for restricted degrees of freedom.
%As mentioned, this is because of the localization of the corner state.

\subsection{System with thermal radiation}\label{s:sb}%---

Finally, we briefly discuss the effect of the thermal radiation. The Stefan-Boltzmann law 
\cinb{gives} rise to nonlinear effect term  in Eq. (\ref{GenDifEqu}). We solve such a nonlinear equation numerically, and apply the 
same procedure in Sec. \ref{s:effective}, which may serve as a mean field approximation.
First of all, let us estimate the order of several parameters of the model.

\begin{figure}[h]
\begin{center}
\begin{tabular}{cc}
\includegraphics[width=0.5\linewidth]{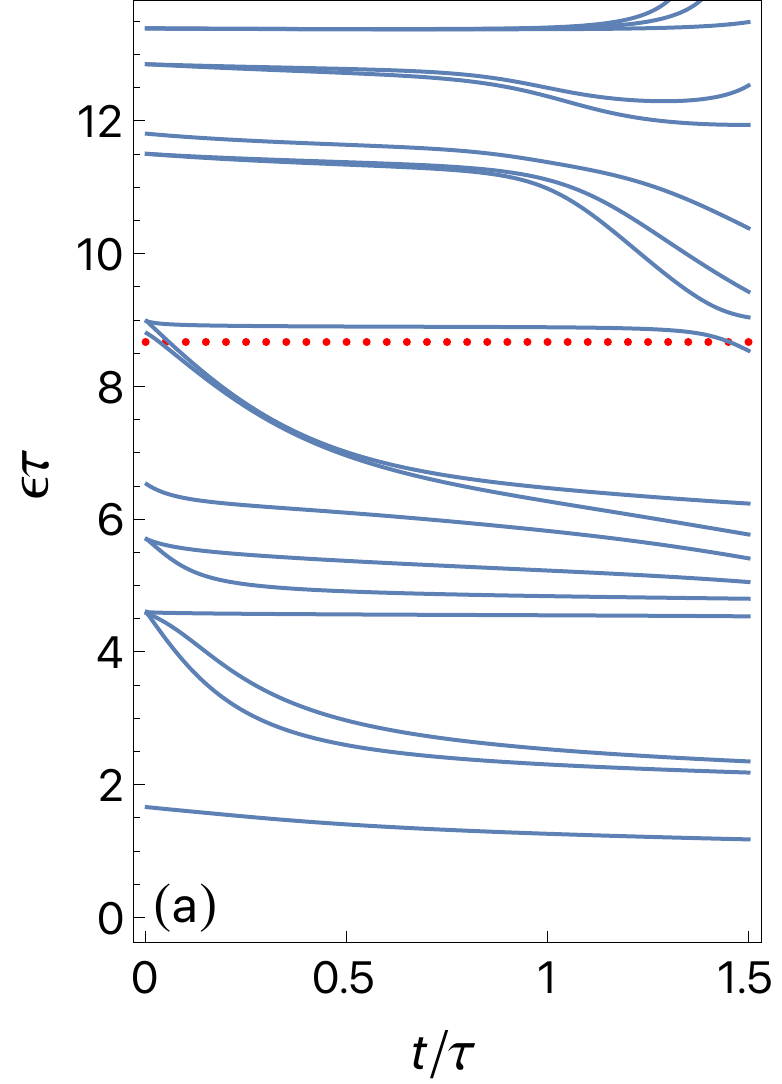}
&\includegraphics[width=0.5\linewidth]{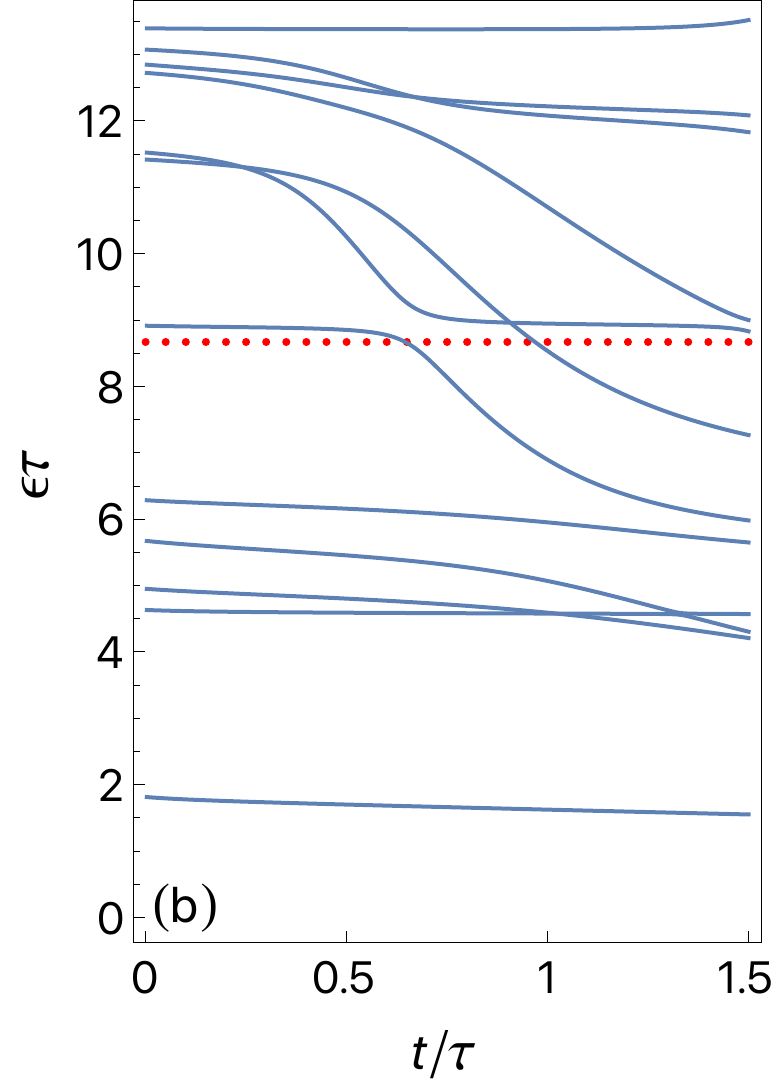}\\
\end{tabular}
\caption{
Same as Fig. \ref{f:eff} but with the Stefan-Boltzmann radiation with $\varepsilon=0.3$.
}
\label{f:thr}%-----------------------------------------------
\end{center}
\end{figure}

The implementation of the experiment in Ref. \cite{Hu:2022aa} is such that  
the bonds have the area of the cross-section $A\sim  10^{-5}\,\hbox{m}^2$ and the length $L\sim115\times10^{-3}\,\hbox{m}$, 
and the nodes have the area of the surface $S\sim \pi R^2 \sim 3\times10^{-5}\,\hbox{m}^2$. 
The room temperature is $T_0=294\,\mbox{K}$ and $\delta T\sim 40\,\mbox{K}$, and hence, $\phi_0\sim7.4$.
From Eqs. (\ref{ParKag1}) and (\ref{ParKag2}), we have
\begin{alignat}1
\frac{\alpha}{D}=\frac{\varepsilon \sigma}{k}\frac{SL}{A}\delta T^3\sim 10^{-10}\varepsilon\delta T^3\sim6.4\varepsilon\times 10^{-5},
\end{alignat}
where we have used $k=200 \hbox{ W/m$\cdot$K}$  for aluminum, and the Stefan-Boltzmann 
constant $\sigma=5.7\times 10^{-8} \hbox{W/m$^2$K$^4$}$
with a relatively large emissivity $\varepsilon=0.3$.
We calculate the energies as follows. Starting from the same initial condition used in Sec. \ref{s:effective}, solving the equations (\ref{HeaEquKag}) numerically, and using data $\bm\phi_p(t)$, 
we compute the eigenvalues of $\cal T$ in Eq. (\ref{CalT}), from which we deduce the energies in the same way in Sec. \ref{s:effective}.
In Fig. \ref{f:thr}, we show the energies thus computed in which small deviation can be seen, but basically, the effect of the thermal radiation is quite small even in a large emissivity. In the realistic case, at most $\varepsilon\sim0.1$, and therefore, 
the thermal radiation may be negligible. 

\section{Summary}
We proposed a higher-order topological heat conduction system toward the detection of the corner states.
We first formulated generic heat conduction on a lattice composed of nodes and bonds, 
based on the Fourier law and energy conservation.
Such a description  indeed has an intimate relationship with the continuum heat conduction equation, as examined 
in the Appendix  using a one-dimensional system with the SSH-like structure.
We next applied to the breathing kagome lattice which is one of typical models showing HOTI phase.
We showed that corner states can be observed by a simple measurement of the temperature only at one corner.
We also proposed more precise measurement of the energy of the corner state using several nodes around a
corner.
Finally, the effect of the Stefan-Boltzmann thermal radiation was estimated, but this was found to be negligible.

\cin{In this paper, we showed that heat conduction on lattices is useful to detect topological properties for corresponding lattice systems. 
We also point out that heat transport from and to reservoirs can also be a useful probe for the same purposes \cite{PhysRevB.96.125144}. 
Thus, 
%it may be interesting to study further applications of heat conduction for the topological problems.
we expect that the heat conduction is much more useful than we expected to reveal directly or indirectly 
topological properties for various systems.
}

\acknowledgements
This work was supported in part by Grants-in-Aid for Scientific Research 
No. 22K03448, No. JP21K13850, and No. JP22H05247
from the Japan Society for the Promotion of Science, and by 
JST CREST, Grant No. JP-MJCR19T1, Japan.

\appendix*
\section{Diffusion on a lattice as a discretization of the diffusion equation}
\label{s:app}

We have introduced diffusion phenomena on a lattice composed of nodes and  bonds.
We show in this Appendix 
that such a description of diffusion phenomena can be regarded as a coarse discretization of the conventional diffusion equation 
if the diffusion constant of a lattice system is interpreted as an effective renormalized parameter.
We show this fact using a simple 1D system with the SSH-like structure.

\begin{figure}[h]
\begin{center}
\begin{tabular}{c}
\includegraphics[width=0.8\linewidth]{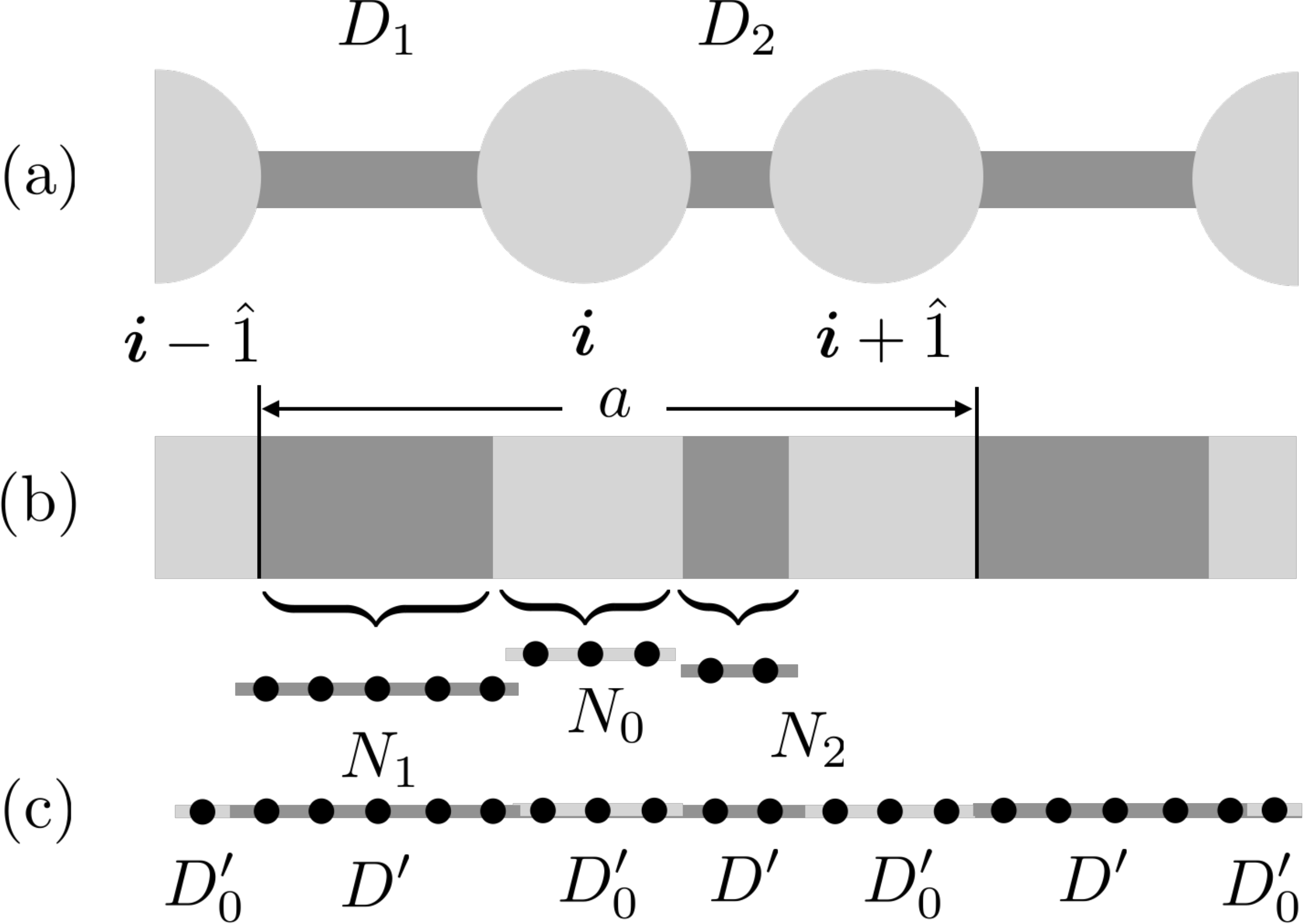}
\end{tabular}
\caption{
Schematic illustration of the 1D system. (a) Experimental implementation in 
\cin{Refs. \cite{https://doi.org/10.1002/adma.202202241,Hu:2022aa}}.
(b) Corresponding continuum system obeying diffusion equation. 
(c) Discretized model of (b), which is numerically solved, instead of (b), in this Appendix.
}
\label{f:1d_lat}%-----------------------------------------------
\end{center}
\end{figure}

%The edge states have been observed in 
The SSH-like structure for the heat conduction \cite{Yoshida:2021vt} has been realized in experimental implementation (a) in Fig. \ref{f:1d_lat} 
\cin{\cite{https://doi.org/10.1002/adma.202202241,Hu:2022aa}},
in which the nodes composed of cylindrical  materials are connected by thin bonds whose
large and small heat conductances are controlled by the length of the bonds.
Apart from the geometric structure, the heat conduction on such a system may be basically described 
by the 1D diffusion equation for type (b) in Fig. \ref{f:1d_lat}, where the geometric structure of (a) is reflected by the 
assumption that the heat capacity of the nodes are very large in (a) and hence, the corresponding parts in (b)  have very small
heat conductivity.
Instead of solving the continuum diffusion equation, we discretized (b) in fine meshes like (c) in Fig. \ref{f:1d_lat}.
In this sense, the heat conduction on lattices in the text is the most coarse discretization 
of the diffusion equation. 

\begin{figure}[h]
\begin{center}
\begin{tabular}{c}
\includegraphics[width=0.7\linewidth]{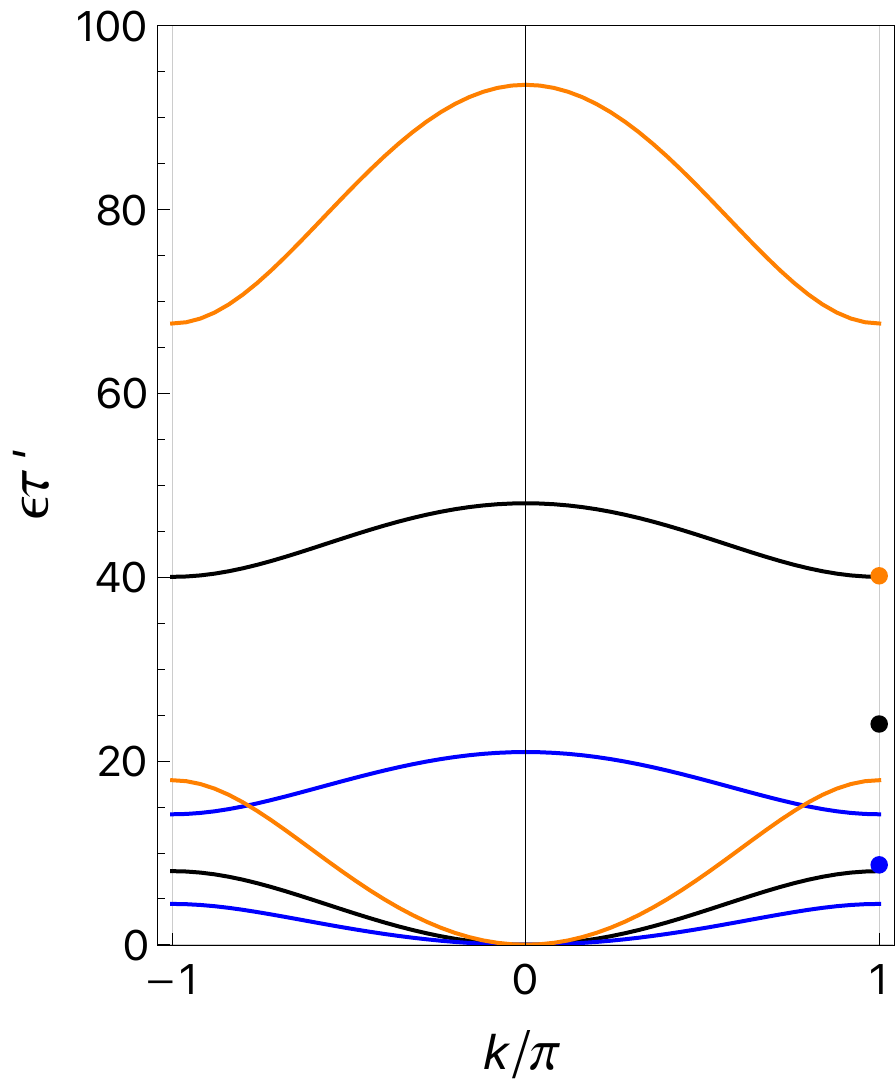}
\end{tabular}
\caption{
The bulk spectra (curves) and edge states (dots). Orange and blue are the cases of
$(N_1,N_2,N_0)=(50,10,5)$, $(20,4,4)$, whereas black is $(0,0,1)$, which is the simple SSH model
in Fig. \ref{f:1d_lat}(a).
%The last one is the SSH model 
The heat conductivities used are $(d,d_0)=(1,10^{-2})$ for colored discretized models, whereas for the SSH model
we have used $(D_1,D_2)=(0.2,1)$. As an effective energy we have used
$1/\tau'=d/(N_1a^2)$ for the discretized models and $D_1/a^2$ for the SSH model, namely the smaller heat 
conductivity of the bonds.
}
\label{f:1d}%-----------------------------------------------
\end{center}
\end{figure}

The discretized Hamiltonian illustrated in Fig. \ref{f:1d_lat}(c) is given by
\begin{alignat}1
H\phi_j&=-\partial^*D'_j\partial\phi_{j}
\nonumber\\&=
-[D'_j\phi_{j+1}+D'_{j-1}\phi_{j-1}-(D'_j+D'_{j-1})\phi_j],
\end{alignat} 
where $\partial $ and $\partial^*$ stand for the forward and backward difference operators, 
$\partial f_j=f_{j+1}-f_j$ and $\partial^* f_j=f_{j}-f_{j-1}$, respectively, and 
\begin{alignat}1
D'_j=\left\{\begin{array}{ll}
\displaystyle D'\equiv d/a_0^2\quad& (j\in\mbox{ bonds})\\
\displaystyle D'_0\equiv d_0/a_0^2  \quad& (j\in\mbox{ nodes})
\end{array}\right. .
\end{alignat}
In the above, $a_0=a/(N_1+N_2+2N_0)$ is the artificial lattice constant of the mesh in (c), where $a$ is the length of the unit cell.
Here we  assume that $d_0\ll d$, since  the heat capacity of the nodes are very large, as mentioned above.
Note here that the bonds have the same heat conductivity $D'$, and the difference of $D_1$ and $D_2$ for the SSH
heat conduction model is due to the difference of the lengths of the bonds, $N_1$ and $N_2$.

In Fig. \ref{f:1d}, we show some examples of the bulk spectra as well as the edge states 
restricted to the two lowest energy bands. 
Here, the edge states are computed by the method proposed in Ref. \cite{Fukui:2020aa}.
The numbers of meshes in a unit cell are $70$ (orange), $34$ (blue), and $2$
(black). The last one corresponds to the SSH heat conduction model
\cite{Yoshida:2021vt,Hu:2022aa}.
One can indeed see that
%that not only the band structure but also the edge states are very similar between very coarse lattice model and 
%small mesh model close to the continuum model.
%Therefore, 
the most coarse lattice model can describe very well the fine-mesh model close to 
the continuum heat conduction equation, including the edge states.
Of course, if one wants to finetune the energies, one should regard $\tau'$ as a parameter, or
 use renormalized diffusion conductivities.
%In the present case, $\tau'$ can be used as an effective parameter connecting continuum model and lattice models.

%\bibliography{kagome_heat}

\end{document}